# *Comparison of Adaptive plan doses using* Velocity® *generated synthetic CT with KV CBCT and re- planning CT*

Sudam Masanta[1*#], Gurvinder Singh[1], Shefali Pahwa[1], Shekhar Dwivedi[1], Devaraju Sampathirao[1], Ramandeep Singh[1]

[1]Department of Radiation Oncology & Medical Physics, Homi Bhabha Cancer Hospital & Research Centre, New Chandigarh, Punjab, India.

*Sudam Masanta to whom any correspondence should be addressed.

E-mail: sudammasanta2017@gmail.com

# Currently working at American Oncology Institute, Jalandhar, Punjab,India.

## Abstract

**Introduction:**

This study uses KV CBCT based Synthetic CT (sCT) generated through Velocity® workstation and compare the target and normal tissue doses with Adaptive plan CT doses.

**Methods:** Thirty head and neck cancer patients undergoing Adaptive Radiation Therapy (ART) were included in this retrospective study. Initially, patient underwent treatment with the primary plan. After subsequent indications of major changes in patients' physicality and anatomy adaptive CT scans were acquired as per institutional protocol. Both the primary planning CT and the indicative cone-beam CT (CBCT) last acquired before the commencement of the adaptive treatment were imported into Velocity® workstation. Rigid and deformable image registration techniques were used for the generation of a Synthetic CT (sCT). Simultaneously replanning was done on re-planning CT (rCT) for adaptive plan execution. The primary plan dose was subsequently mapped and deformed onto the Synthetic CT in Velocity® workstation, allowing for a comparative dosimetric analysis between the sCT and rCT plan doses. This comparison was conducted in both Velocity® and Eclipse, focusing on dose variations across different organs at risk (OARs) and the planning target volume (PTV). Additionally, dosimetric indices were evaluated to assess and validate the accuracy and quality of the synthetic CT-based dose mapping relative to adaptive planning.

**Results:** The dosimetric comparison between sCT and rCT stated that Mean dose for OARs and PTVs were found to be similar in the two planning and the level of confidence by using T-statistics. Collaborative research has the potential to eliminate the need of rCT as a standard requirement.

**Conclusion:** The sCT shows comparable CT numbers and doses to the replanning CT, suggesting it's potential as a replacement pending clinical corelation and contour adjustments.

# 1. Introduction

Imaging is an indispensable component of radiotherapy treatment, ensuring precise diagnosis, simulation, treatment planning, and real-time patient anatomy verification through kilovoltage cone-beam CT (KV-CBCT). Advanced imaging techniques are fundamental for delivering high-precision treatment in Volumetric Modulated Arc Therapy (VMAT), Three-Dimensional Conformal Radiotherapy (3DCRT), and Intensity-Modulated Radiotherapy (IMRT).

Radiotherapy remains a cornerstone of cancer treatment, where accurate radiation delivery to the tumor is paramount for achieving optimal clinical outcomes. Traditionally, treatment planning has relied on pre-treatment imaging under the assumption that tumor shape, position, and surrounding normal tissues remain constant throughout the whole course of treatment. However, in reality, tumors undergo shrinkage, displacement, and anatomical alterations, while normal tissues dynamically respond to radiation. These unpredictable variations compromise dose accuracy, potentially reducing tumor control and increasing the risk of radiation-induced toxicity to healthy tissue [1-3,4-7].

To combat these challenges, Adaptive Radiotherapy (ART) has revolutionized cancer treatment by dynamically modifying treatment plans in response to periodic imaging, typically CBCT or MRI. ART enables real-time adaptation to fluctuations in tumor volume, position, and organ motion, ensuring optimal radiation precision while significantly minimizing damage to healthy tissues. Unlike conventional static treatment plans, ART enhances dose accuracy and treatment efficacy, mitigating the risk of suboptimal dose distribution [8-10].

Modern linear accelerators are now equipped with CBCT technology, allowing for precise patient setup verification and continuous anatomical monitoring during therapy. CBCT is particularly crucial for head-and-neck cancer treatments, where anatomical changes are frequent and substantial. While CBCT-based planning has been explored, scatter and noise artifacts continue to hinder image quality and dosimetric accuracy, posing challenges for treatment precision [11-13].

To overcome these limitations, rigid and deformable image registration-based synthetic CT has emerged as an alternative solution for treatment planning [14-16]. In this study, Velocity® was employed for a dosimetric comparison between two plans. The synthetic CT generated in Velocity® seamlessly integrates deformed structures, including organs at risk (OARs) and planning target volumes (PTVs). To ensure accurate evaluation, both adapted and primary fraction doses were mapped onto the synthetic CT, enabling a comprehensive dosimetric assessment.

This study aims to validate the feasibility, accuracy, and clinical advantages of synthetic CT-based planning in radiotherapy. By addressing critical challenges such as noise and scatter, synthetic CT has the potential to enhance treatment precision, improve patient outcomes, and redefine the future of adaptive radiotherapy.

# 2. Methodology

## 2.1 Patient Selection

This study is a purely retrospective dosimetric comparison involving 30 patients with different planning target volume (PTV) doses and varying tumor volumes. The selected patients underwent adaptive radiotherapy (ART) between 2022 and 2024. CT images for treatment planning were acquired using a SOMATOM Confidence Siemens Computed Tomography with an 80 cm bore diameter. Patients included in this study received Volumetric Modulated Arc Therapy (VMAT) with prescribed doses of 66Gy, and 60Gy in 30 fractions and 70 Gy in 35 fractions over 6 to 7 weeks. All treatment planning was conducted in the Eclipse v 16.1 Treatment Planning System (TPS) using 6 MV photon energy.

## 2.2 CBCT in Synthetic CT generation

Modern Varian linear accelerators are equipped with an On-Board Imager (OBI) for kilovoltage cone-beam CT (KV-CBCT), which is used for daily anatomical verification and patient setup accuracy.

For head-and-neck patients, CBCT imaging was performed at 100 kVp and 150 mAs. When significant anatomical changes were observed, patient was re-simulated & new planning CT was acquired which was imported to Eclipse for re-planning. Later on, this was imported into Velocity® and a synthetic CT (sCT) was then generated using rigid and deformable image registration between the primary CT and indicative KV-CBCT. The structure set from the primary planning CT was mapped onto the synthetic CT using deformable registration, ensuring anatomical consistency for dosimetric evaluation.

## 2.3 Image Verification

The primary goal of image verification is to evaluate Hounsfield Unit (HU) accuracy, as changes in HU values can impact dose distribution. In head-and-neck cases, different regions exhibit varying HU values, influencing dose calculations. In literature, studies shows that the standard deviation of HU for different organs was analysed and compared between synthetic CT and primary CT to assess image quality and accuracy [17,18].

## 2.4 Dosimetric Parameters for evaluation

To ensure a robust comparison, identical dosimetric parameters were devised for both synthetic CT-based adaptive plans and rCT based adaptive planning in total fraction-based plans in both Velocity® and eclipse workstation respectively. From the dose-volume histogram (DVH), the following metrics were extracted:

a) $V_{PTV}$ (Volume of the Planning Target Volume)
b) $V_{95\%}$ (volume receiving 95% of the prescribed dose.)
c) $D_2\%$ (dose received by the 2% of the volume of the PTV.)
d) $D_{98}\%$ (dose received by the 98% of the volume of the PTV.)
e) $D_5\%$ (dose received by the 5% of the volume of the PTV.)
f) $D_{95}\%$ (dose received by the 95% of the volume of the PTV.)

Additionally, the confidence level of probability was calculated for various organs at risk (OARs), including:

a) Left Parotid
b) Right Parotid
c) Oral Cavity
d) Larynx
e) Spinal Cord
f) Thyroid

This methodology ensures a comprehensive evaluation of the dosimetric impact of synthetic CT-based adaptive planning compared to replanning-based adaptive treatment.

### 2.5 Dosimetric Indices

Dosimetric indices are mathematical and graphical tools used to analyze and compare radiotherapy plans. Throughout the evolutionary trajectory of radiotherapy, characterized by successive advancements in technology and the continual refinement of irradiation methodologies, the overarching objective has remained the precise and homogeneous deposition of the entirety of the prescribed radiation dose across the totality of the delineated target volume. This encompasses both the macroscopically identifiable tumor mass and the microscopic tumor cells potentially residing within the designated treatment region. Simultaneously, paramount emphasis has been placed on the stringent minimization of radiation exposure to the contiguous healthy tissues, thereby optimizing the therapeutic ratio and mitigating undue iatrogenic effects. In 1993, Radiation Therapy Oncology Group (RTOG) propose Conformity Index firstly and it also described in ICRU-62 report. The Conformity Index (CI) emerged as a sophisticated extension of sectional dosimetric assessments and dose–volume histograms, serving as a quantitative metric to evaluate the precision of radiation dose distribution. It can be rigorously defined as an absolute parameter derived from the correlation between the delineated tumor volume, or its fractional representation, and the volumetric extent encompassed by a specific isodose threshold or its proportional counterpart. Furthermore, CI may be delineated as the ratio between distinct isodose levels, encompassing the prescription isodose, reference isodose, minimum isodose, and maximum isodose, thereby enabling a nuanced evaluation of dose conformity [19]. The implementation of this quantitative analytical tool is instrumental in optimizing treatment strategy selection and facilitating comparative assessments across diverse radiotherapeutic modalities, including conformal radiotherapy, stereotactic radiotherapy, and brachytherapy. This approach aligns with the recommendations set forth by the American Brachytherapy Society, reinforcing its clinical utility in treatment planning and quality assurance.

$$\textbf{Conformity Index (RTOG)} = \frac{V_{RI}}{V_{PTV}}$$

where $V_{RI}$ volume of the reference isodose, and $V_{PTV}$ is the volume of the PTV.

The Homogeneity Index (HI) is a critical dosimetric parameter that is inherently influenced by multiple factors, including target volume, anatomical location, and prescribed radiation dose.

The degree of tissue heterogeneity varies significantly across different anatomical regions of the body, directly impacting dose distribution. Among all sites, the brain exhibits the least heterogeneity due to minimal density variations, in stark contrast to the head and neck, thoracic, abdominal, and pelvic regions. The head and neck region demonstrates the highest degree of density heterogeneity, primarily due to complex anatomical structures such as the oral cavity, nasal cavity, high-density bone, teeth, tongue, and, in certain cases, dental implants. These factors introduce substantial perturbations in radiation dose distribution within the target volume, necessitating meticulous treatment planning.

Notably, clinical observations consistently indicate that treatment plans for intracranial cases yield superior homogeneity within the Planned Target Volume (PTV), ensuring a more uniform and controlled dose deposition. In this study we are including only ICRU-83 based Homogeneity Index [20].

Mathematical formulas of homogeneity indices published in literature [20].

| Author | Formulation | Parameter Description |
|---|---|---|
| RTOG | $HI_{RTOG} = \dfrac{I_{max}}{RI}$ | $I_{max}$ & RI are the maximum dose and reference dose to PTV. |
| RTOG | $HI = \dfrac{D_5}{D_{95}}$ | $D_5$, $D_{95}$ are the doses to 5% & 95% volume of the PTV. |
| ICRU-62 | $HI = \dfrac{D_{max}}{D_{min}}$ | $D_{max}$ & $D_{min}$ are the maximum and minimum dose in PTV. |
| ICRU-83 | $HI = \dfrac{D_2 - D_{98}}{D_P} \times 100$ | $D_2$ & $D_{98}$ are the doses to 2% & 98% volume of the PTV. |
| ICRU-83 | $HI = \dfrac{D_5 - D_{95}}{D_P} \times 100$ | $D_5$, $D_{95}$ are the doses to 5% & 95% volume of the PTV. |
| ICRU-62 | $HI = \dfrac{D_{max}}{D_P}$ | $D_{max}$ & $D_P$ are the maximum and prescribed dose in PTV. |
| ICRU-83 | $HI = \dfrac{D_5 - D_{95}}{D_{50}} \times 100$ | $D_5$, $D_{95}$, $D_{50}$ are the doses to 5% & 95%, 50% volume of the PTV |

## 3. Results

### 3.1 Volume of PTVs

Table 1 presents the mean and standard deviation of $V_{95\%}$ and $V_{PTV}$. The percentage variation in the mean values between $V_{95\%}$ and $V_{PTV}$ of synthetic CT (sCT) and replanning CT (rCT) is found to be -2.51 % and -5.41% respectively. The negative values in the percentage difference indicate an increase in volumetric coverage in the synthetic CT compared to the adaptive CT. The statistical analysis reveals a P-value for $V_{95\%}$ and $V_{PTV}$ of 0.449 and 0.376 respectively, corroborating the lack of statistical significance in volumetric disparity.

**Table 1.** Mean and Standard deviation of PTVs

| Variable | Mean (cc) | | P value | % Mean |
|---|---|---|---|---|
| | sCT | rCT | | |
| $V_{95}$ | 307.8051 | 301.307 | 0.449>0.005 | -2.51 |
| $V_{PTV}$ | 323.88 | 307.251 | 0.376>0.005 | -5.41 |

### 3.2 DVH parameters of different PTVs doses

Table 2 provides a detailed comparison of the mean and standard deviation values for the dose parameters D2%, D98%, D95%, and D5% of 60Gy in 30 Fractions. The maximum percentage variation observed among these dose values is 0.38%. Furthermore, the results of the statistical analysis indicate that the P-value is greater than 0.05. Since a P-value above 0.05 suggests a lack of statistically significant differences, it can be concluded that there is no meaningful variation between the two types of CT scans being analysed, namely synthetic CT (sCT) and replanning CT (rCT). Table 3 provides a detailed comparison of the mean and standard deviation values for the dose parameters $D_2\%$, $D_{98}\%$, $D_{95}\%$, and $D_5\%$ of 66Gy in 30 Fractions. Table 4 provides a detailed comparison of the mean and standard deviation values for the dose parameters $D_2\%$, $D_{98}\%$, $D_{95}\%$, and $D_5\%$ of 70Gy in 35 Fractions.

**Table 2.** Mean and standard deviation of dose of PTV (60Gy in 30Fr).

| DVH Parameters | Mean (cGy) | | P Value | % Mean |
|---|---|---|---|---|
| | sCT | rCT | | |
| $D_2\%$ | 6211.769 | 6228.015 | 0.224 | 0.27 |
| $D_{98}\%$ | 5724.615 | 5746.254 | 0.1283 | 0.38 |
| $D_5\%$ | 6182.154 | 6187.077 | 0.3935 | 0.08 |
| $D_{95}\%$ | 5814.846 | 5831.123 | 0.1462 | 0.28 |

**Table 3.** Mean and standard deviation of dose of PTV (66Gy in 30Fr).

| DVH Parameters | Mean (cGy) | | P Value | % Mean |
|---|---|---|---|---|
| | sCT | rCT | | |
| $D_2\%$ | 6712.89 | 6776.6 | 0.218 | 0.94 |
| $D_{98}\%$ | 6165.6 | 6157.11 | 0.473 | 0.18 |
| $D_5\%$ | 6743.2 | 6746.05 | 0.4861 | 0.042 |
| $D_{95}\%$ | 6322 | 6361.69 | 0.2152 | -0.62 |

**Table 4.** Mean and standard deviation of dose of PTV (70Gy in 35Fr).

| DVH Parameters | Mean(cGy) | | P Value | % Mean |
|---|---|---|---|---|
| | sCT | rCT | | |
| $D_2\%$ | 7187.2 | 7265.04 | 0.09 | -1.07 |
| $D_{98}\%$ | 6672.2 | 6720.32 | 0.186 | 0.71 |
| $D_5\%$ | 7157.8 | 7176.2 | 0.207 | -0.26 |
| $D_{95}\%$ | 6770.2 | 6807.88 | 0.2045 | -0.55 |

**Fig1:** Mean Dose Distribution to PTV in sCT vs rCT(60Gy in 30 Fr)

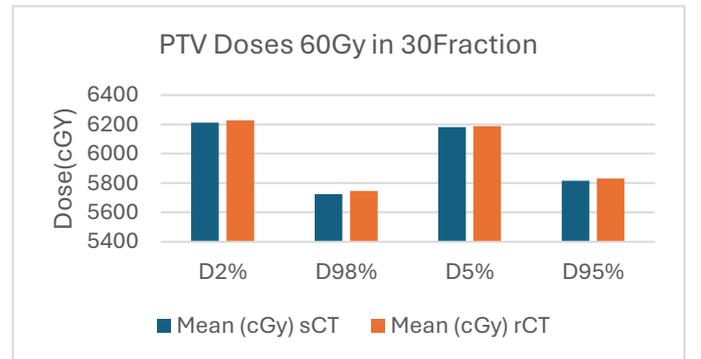

**Fig2:** Mean Dose Distribution to PTV in sCT vs rCT(66Gy in 30Fr)

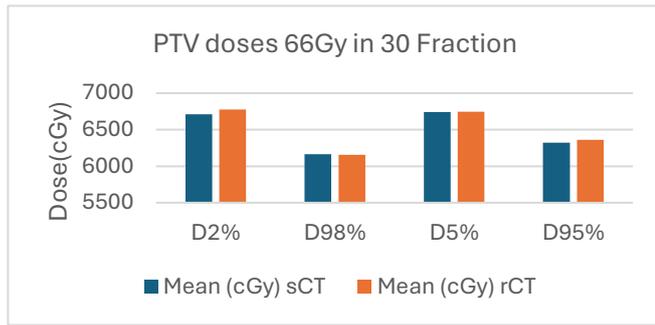

**Fig3:** Mean Dose Distribution to PTV in sCT vs rCT (70Gy in 35Fr)

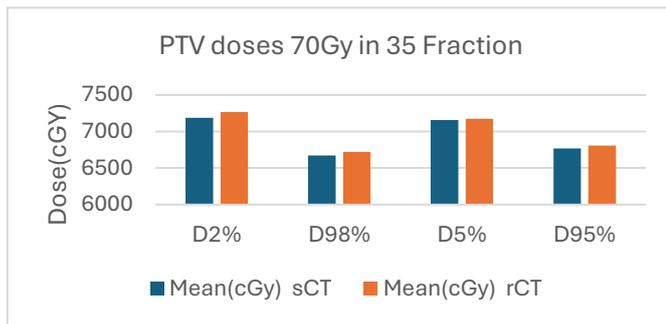

The above bar chart illustrates the comparison of Planning Target Volume (PTV) dose parameters between the original sCT and the rCT plans. The dose metrics evaluated include $D_2\%$, $D_{98}\%$, $D_5\%$, and $D_{95}\%$, with dose values expressed in 60Gy in 30Fraction and 66 Gy in 30 Fraction, 70 Gy in 35 Fraction. Each parameter is represented with blue bars for sCT and orange bars for rCT. Overall, the rCT dose distribution appears to be slightly elevated across all PTV dose metrics, suggesting that the adaptive plan may provide improved target coverage or compensate for anatomical or positional changes during the treatment course. The sCT dose is shown in fig.4 in Velocity® workstation.

**Fig 4.** sCT dose in Velocity® Workstation

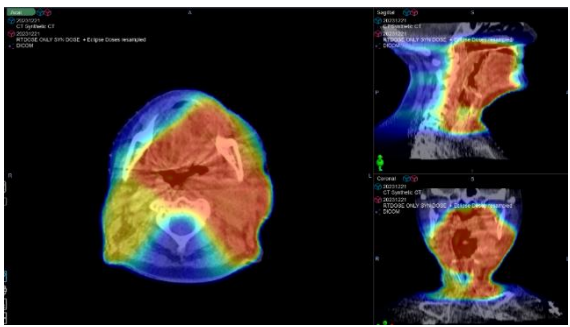

### 3.3 DVH parameters of different OARs

Table 5, shows the Dose Volume histogram (DVH) for different organs (OARs). From this table it is observed that,

**Table 5.** Mean and P Value of OARs (70Gy in 35 Fr)

|  | OAR | Mean | | P value |
|---|---|---|---|---|
|  |  | sCT(cGy) | rCT(cGy) |  |
| $D_{Mean}$ | Parotid -L | 3435.6 | 3593.5 | 0.45 |
|  | Parotid -R | 2508.8 | 2733.97 | 0.398 |
|  | Oral Cavity | 3292 | 4171 | 0.28 |
|  | Thyroid | 4546.4 | 4548.086 | 0.499 |
|  | Larynx | 4816.6 | 5022.38 | 0.437 |
| $D_{Max}$ | Spinal Cord | 3470 | 3465.95 | 0.494 |
|  | Mandible | 6152.6 | 7299.7 | 0.147 |

**Table 6.** Mean and P Value of OARs (66Gy in 30 Fr)

|  | OAR | Mean | | P value |
|---|---|---|---|---|
|  |  | sCT(cGy) | rCT(cGy) |  |
| $D_{Mean}$ | Parotid -L | 2179.7 | 2016.28 | 0.33 |
|  | Parotid -R | 3274 | 2994.2 | 0.322 |
|  | Oral Cavity | 4049.5 | 4027.5 | 0.489 |
|  | Thyroid | 5201.7 | 5514.68 | 0.205 |
|  | Larynx | 4895.9 | 4692.6 | 0.357 |
| $D_{Max}$ | Spinal Cord | 3538.6 | 3473.42 | 0.343 |
|  | Mandible | 6379 | 6533 | 0.357 |

**Table 7.** Mean and P Value of OARs (60Gy in 30 Fr)

|  | OAR | Mean | | P value |
|---|---|---|---|---|
|  |  | sCT(cGy) | rCT(cGy) |  |
| $D_{Mean}$ | Parotid -L | 2751.25 | 2668.59 | 0.457 |
|  | Parotid -R | 2989.23 | 3102.15 | 0.441 |
|  | Oral Cavity | 4528.5 | 4387 | 0.397 |
|  | Thyroid | 4069.75 | 4097.47 | 0.484 |
|  | Larynx | 3313.62 | 3038.59 | 0.308 |
| $D_{Max}$ | Spinal Cord | 2957.77 | 2945 | 0.482 |
|  | Mandible | 6119.78 | 6186.67 | 0.337 |

**Fig5:** Mean Dose Distribution to OARs in sCT vs rCT (66Gy in 30Fr)

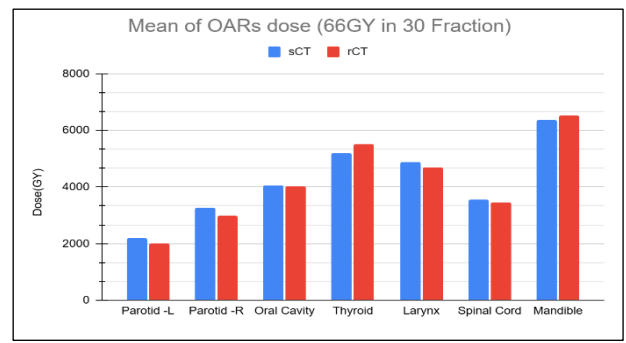

**Fig6:** Mean Dose Distribution to OARs in sCT vs rCT (70Gy in 35Fr)

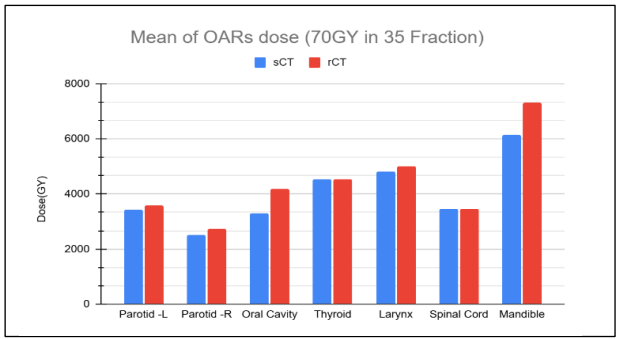

**Fig7:** Mean Dose Distribution to OARs in sCT vs rCT (60Gy in 30Fr)

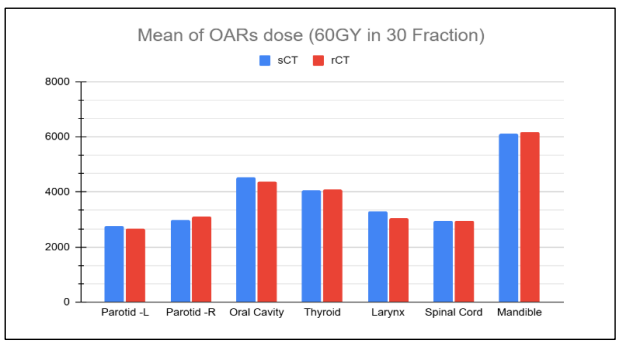

The mean doses (in cGy) delivered to various organs at risk (OARs) in the sCT and re-CT (rCT) plans are illustrated in Figure 5. As shown, there are notable variations in the dose distributions between the two plans. For instance, the mean dose to the left parotid gland (Parotid-L) increased slightly in the rCT compared to the sCT plan. The right parotid gland (Parotid-R) showed a marginal increase as well. These changes emphasize the importance of adaptive planning to minimize unnecessary exposure to OARs.

### 3.4 Dosimetric Parameters

Table 8,9,10 present a comparison of key dosimetric indices. In particular, the variation in the Conformity Index between the two types of CT scans is analysed. The percentage of mean variation between the synthetic CT (sCT) and reference CT (rCT) is found to be -3.8% and -3.06%, -3.67% for 70Gy in 35 fractions, 66GY in 30 Fractions and 60GY in 30 Fractions respectively. The negative value indicates that the volume receiving 95% of the prescribed isodose is smaller than the Planned Target Volume (PTV), suggesting a potential under coverage of the target region.

**Table 8** Mean and standard deviation of CI (70Gy in 35 Fr)

| Dosimetric Parameters | Mean | | P Value | % Mean |
|---|---|---|---|---|
| | sCT | rCT | | |
| CI | 0.9517 | 0.9901 | <0.001 | -3.8 |

**Table 9** Mean and standard deviation of CI(66 Gy in 30Fr)

| Dosimetric Parameters | Mean | | P Value | % Mean |
|---|---|---|---|---|
| | sCT | rCT | | |
| CI | 0.95 | 0.98 | <0.001 | -3.06 |

**Table 10** Mean and standard deviation of CI(60 Gy in 30Fr)

| Dosimetric Parameters | Mean | | P Value | % Mean |
|---|---|---|---|---|
| | sCT | rCT | | |
| CI | 0.95 | 0.9862 | <0.001 | -3.67 |

Similarly, Table 11,12,13 highlights another significant dosimetric index that has been analysed and compared between sCT and rCT. Specifically, the Homogeneity Index (HI) is calculated using different guidelines, including those established by the Radiation Therapy Oncology Group (RTOG) and various formulations outlined in the International Commission on Radiation Units and Measurements (ICRU) Report No. 83 & 62. These comparisons provide insights into the consistency of dose distribution within the target volume across different calculation methods [21].

**Table 11** Mean and standard deviation of CI(70 Gy in 35Fr)

| HI | sCT | rCT | % Mean |
|---|---|---|---|
| $D_5\%/D_{95}\%$ | 1.0573 | 1.0541 | 0.3 |
| $(D_{2\%} - D_{98\%})/D_P$ | 0.074 | 0.077 | -3.89 |
| $D_{Max}/D_P$ | 1.0544 | 1.0672 | 1.21 |

**Table 12** Mean and standard deviation of CI(60 Gy in 30Fr)

| HI | sCT | rCT | % Mean |
|---|---|---|---|
| $D_5\%/D_{95}\%$ | 1.0632 | 1.0619 | -0.12 |
| $(D_{2\%} - D_{98\%})/D_P$ | 0.084 | 0.08029 | -4.6 |
| $D_{Max}/D_P$ | 1.0683 | 1.0758 | 0.65 |

**Table 13** Mean and standard deviation of CI(66 Gy in 30Fr)

| HI | sCT | rCT | % Mean |
|---|---|---|---|
| $D_5\%/D_{95}\%$ | 1.0632 | 1.0619 | -0.12 |
| $(D_{2\%} - D_{98}\%)/D_P$ | 0.084 | 0.08029 | -4.6 |
| $D_{Max}/D_P$ | 1.0683 | 1.0758 | 0.65 |

In some cases, the PTV does not exactly match with the rCT PTV, resulting in changes in PTV doses with respect to the rCT PTV doses. For OARs, the $D_{Max}$ doses can change rapidly even with slight changes in the OAR contours. In organs like the spinal cord, significant dose variations have been observed in some patients when there is a positional mismatch. In some studies, they observed the same issue [17]. Therefore, in certain cases, replacing the rCT procedures might lead to improved results.

## 4. Discussion

This study aimed to evaluate the use of synthetic CT (sCT), generated via deformable image registration (DIR) of primary CT (pCT) and indicative CBCT, for predicting the dose delivered to patients, thereby supporting clinical decisions regarding the need for adaptive radiotherapy (ART). Determining whether ART is necessary requires the establishment of a threshold beyond which anatomical or dosimetric changes warrant adaptation of the treatment plan [22]. Another objective of this study was to assess whether sCT could effectively replace the need for re-planning CT (rCT), potentially streamlining the treatment workflow. Utilizing sCT derived from CBCT for adaptive planning may reduce delays and help prevent interruptions in radiotherapy—particularly during the critical fourth and fifth treatment weeks, which can adversely affect tumor control probability. This approach may also mitigate risks such as skin toxicity caused by mid-treatment anatomical changes. Overall, the integration of sCT in adaptive workflows offers the potential to conserve resources, minimize radiation exposure, and enhance treatment efficiency and patient safety. However, prior studies using various DIR algorithms in the pelvic region have shown that accurate deformation of large volumes, such as the bladder, remains challenging—an observation that aligns with the findings of our study [1,2]. Specifically, lower D95 and D98 values observed in sCT compared to rCT can be partially attributed to discrepancies in bladder volume deformation. In earlier research, CBCT and rCT scans were acquired 5–7 days apart, leading to significant anatomical and positional changes. In an effort to compensate, the CBCT was first deformed to match the rCT, followed by deforming the pCT to the adjusted CBCT, thereby improving anatomical consistency—but introducing unquantified uncertainties due to compounded registration errors [23]. To reduce such uncertainty, our study maintained a shorter time gap (2–3 days) between CBCT and rCT acquisition, which was the best compromise possible given our department's high patient volume and limited simulation slot availability. Ideally, both scans should be taken on the same day with consistent patient positioning and immobilization, though practical constraints often prevent this. Despite these limitations, our results showed no major differences in $D_5\%$, $D_{95}\%$, $D_2\%$, and $D_{98}\%$ for the planning target volume (PTV), indicating generally reliable dose estimations. While this study did not focus on evaluating contour propagation accuracy, all sCT contours were initially generated through automatic propagation and required only minimal manual adjustments. One of the key limitations of this work is the residual anatomical and positional mismatches between CBCT and rCT scans, despite careful efforts to replicate patient positioning. These mismatches may have influenced dose evaluation results and could be especially significant for OARs like the spinal cord, where clinical endpoints (e.g., risk of radiation myelopathy) are determined based on maximum or near-maximum dose values. Even small discrepancies in alignment can result in meaningful differences in these critical metrics. To address this, future work should aim to eliminate positional variability, potentially by acquiring both CBCT and rCT on the same day using identical immobilization setups. While implementing such a protocol is challenging in busy clinical environments, it could significantly improve the precision and reliability of adaptive radiotherapy workflows.

## Author Contributions

All authors have substantially contributed to the conception and design of the study, acquisition, analysis, and interpretation of data.

## Declaration of Generative AI and AI-assisted technologies in the writing process

During the study of this work the author(s) used Grammarly in order to correct for grammatical errors in the writing process. After using this tool/service, the author(s) reviewed and edited the content as needed and take(s) full responsibility for the content of the publication.

## Acknowledgements

We would like to thank the entire Radiation Oncology and medical physics Department for the consistent support and inputs in carrying out the study.

## References

[1] Kataria T, Gupta D, Goyal S, Bisht SS, Abhishek A, Subramani V, et al. Clinical outcomes of adaptive radiotherapy in head and neck cancers. Br J Radiol. 89 (1062):20160085. https://doi.org/10.1259/bjr.20160085.

[2] Surucu M, Shah KK, Roeske JC, Choi M, Small Jr W, Emami B, et al. Adaptive radiotherapy for head and neck cancer. Technol Cancer Res Treat 2017;16(2): 218–23. https://doi.org/10.1177/1533034616662165.

[3] Gensheimer MF, Le QT. Adaptive radiotherapy for head and neck cancer: are we ready to put it into routine clinical practice? Oral Oncol 2018;86:19–24. https://doi.org/10.1016/j.oraloncology.2018.08.010.

[4] Castelli J, Simon A, Louvel G, Haigron P, Acosta O, Rigaud B, et al. Impact of head and neck cancer adaptive radiotherapy to spare the parotid glands and decrease the risk of xerostomia. Radiat Oncol Lond Engl 2015;10:6. https://doi.org/10.1016/j.oraloncology.2018.08.010.

[5] Hansen EK, Bucci MK, Quivey JM, Verma V, Mayr NA, McSherry F, et al. Repeat CT imaging and replanning during the course of IMRT for head-and-neck cancer. Int J Radiat Oncol 2006;64(2):355–62. https://doi.org/10.1016/j.ijrobp.2005.07.957.

[6] Yang H, Hu W, Wang W, Zhang X, Li Y, Zhao W, et al. Replanning during intensity modulated radiation therapy improved quality of life in patients with nasopharyngeal carcinoma. Int J Radiat Oncol Biol Phys 2013;85(1):e47–54. https://doi.org/10.1016/j.ijrobp.2012.09.033.

[7] Barateau A, Garlopeau C, Cugny A, Lemoine J, Soler L, Reniers B, et al. Dose calculation accuracy of different image value to density tables for cone-beam CT planning in head & neck and pelvic localizations. Phys Medica PM Int J Devoted Appl Phys Med Biol Off J Ital Assoc Biomed Phys AIFB 2015;31(2):146–51. https://doi.org/10.1016/j.ijrobp.2012.09.033

[8] Rigaud B, Simon A, Castelli J, Lafond C, Haigron P, Acosta O, et al. Deformable image registration for radiation therapy: principle, methods, applications, and evaluation. Acta Oncol Stockh Swed 2019;58(9):1225–37. https://doi.org/10.1080/0284186X.2019.1620331.

[9] Oh S, Kim S. Deformable image registration in radiation therapy. Radiat Oncol J 2017;35(2):101–11. https://doi.org/10.3857/roj.2017.00325.

[10] Castelli J, Simon A, Lafond C, Rigaud B, Haigron P, Acosta O, et al. Adaptive radiotherapy for head and neck cancer. Acta Oncol Stockh Swed 2018;57(10): 1284–92. https://doi.org/10.1080/0284186X.2018.1505053.

[11] Richter A, Hu Q, Steglich D, Müller J, Thomas S, Albrecht T, et al. Investigation of the usability of cone-beam CT data sets for dose calculation. Radiat Oncol Lond Engl 2008;3:42. https://doi.org/10.1186/1748-717X-3-42.

[12] Marchant TE, Joshi KD, Moore CJ. Shading correction for cone-beam CT in radiotherapy: validation of dose calculation accuracy using clinical images. In: Medical Imaging 2017: Physics of Medical Imaging. Vol 10132. SPIE; 2017:119- 129. https://doi.org/10.1117/12.2254059.

[13] Varian Medical Systems. Velocity 4.0 Instructions for Use. https://www.myvarian.com/s/productdocumentationdetail?Id=06944000003dEYSAA2%26lang=en; 2024 [accessed 14 May 2024].

[14] Fox T, Andl G, Bose S. Velocity and Deformable Image Registration. https://varian.widen.net/view/pdf/dizhixzecz/VelocityClinicalPerspectives_RAD10438_February2018.pdf?u=bmxzem; 2018 [accessed 1 January 2024].

[15] Loeckx D, Maes F, Vandermeulen D, Defraene G, Sima DM, Bogaerts R, et al. Nonrigid image registration using a statistical spline deformation model. Inf Process Med Imaging Proc Conf 2003;18:463–74. https://doi.org/10.1007/978-3-540-45087-0_39.

[16] Lawson JD, Schreibmann E, Jani AB, Dong L, Yang Y, Hagan M, et al. Quantitative evaluation of a cone-beam computed tomography–planning computed tomography deformable image registration method for adaptive radiation therapy. J Appl Clin Med Phys 2007;8(4):96–113. https://doi.org/10.1120/jacmp.v8i4.2432.

[17] Assessment of the use of synthetic CT produced by deformable image registration of planning CT and CBCT in adaptive radiotherapy treatments of head and neck cancers. https://doi.org/10.1016/j.ejmp.2025.104929

[18] Bissonnette JP, Balter PA, Dong L, Starkschall G, Dunscombe P, Oldham M, et al. Quality assurance for image-guided radiation therapy utilizing CT-based technologies: a report of the AAPM TG-179. Med Phys 2012;39(4):1946–63. https://doi.org/10.1118/1.3690466

[19] Conformity index: A review- https://doi.org/10.1016/j.ijrobp.2005.09.028.

[20] Plan evaluation indices: A journey of evolution. https://doi.org/10.1016/j.rpor.2020.03.002

[21] A new homogeneity index definition for evaluation of radiotherapy plans. https://aapm.onlinelibrary.wiley.com/share/SPVF4I54C7WUCJ7FVE4D?target=10.1002/acm2.12739.

[22] Zou KH, Warfield SK, Bharatha A, Kaus M, Rusinek H, Grimson W, et al. Statistical validation of image segmentation quality based on a spatial overlap index. Acad Radiol 2004;11(2):178–89. https://doi.org/10.1016/s1076-6332(03)00671-8.

[23] Chang Y, Liang Y, Yang B, Zhang X, Li Y, Chen Y, et al. Dosimetric comparison of deformable image registration and synthetic CT generation based on CBCT images for organs at risk in cervical cancer radiotherapy. Radiat Oncol Lond Engl 2023;18(1):3. https://doi.org/10.1186/s13014-022-02191-3.